\documentclass[reprint,amsmath,amssymb,apr,aip,onecolumn]{revtex4-2}

\usepackage{graphicx}
\usepackage{graphics}
\usepackage{algorithm,algorithmic}
\usepackage{wrapfig}
\usepackage{mathptmx}
\usepackage{times}
\usepackage{amsmath}
\usepackage{amssymb}
\usepackage{dcolumn}
\usepackage{color}
\usepackage{bm}
\usepackage{units}
\usepackage{booktabs}
\usepackage{multirow}
\usepackage[version=4]{mhchem}
\usepackage{siunitx}[=v2]

\begin{document}

\title{Emerging Trends in Intelligent Sensing}

\author{Ghazi Sarwat Syed}\affiliation{IBM Research -- Europe, S\"{a}umerstrasse 4, 8803 R\"{u}schlikon, Switzerland}\email{ghs@zurich.ibm.com}
\maketitle

\noindent\textbf{The rapid proliferation of artificial intelligence, connected devices, and high speed mobile networks is driving unprecedented computational demands that challenge traditional sensor architectures. This article explores the shift toward edge computing, where computation is performed directly at the data source, and highlights the key architectures and performance metrics that may define the next generation of intelligent sensor systems.} \\



\begin{wrapfigure}{r}{0.56\textwidth}
    \centering
    \includegraphics[width=0.53\textwidth]{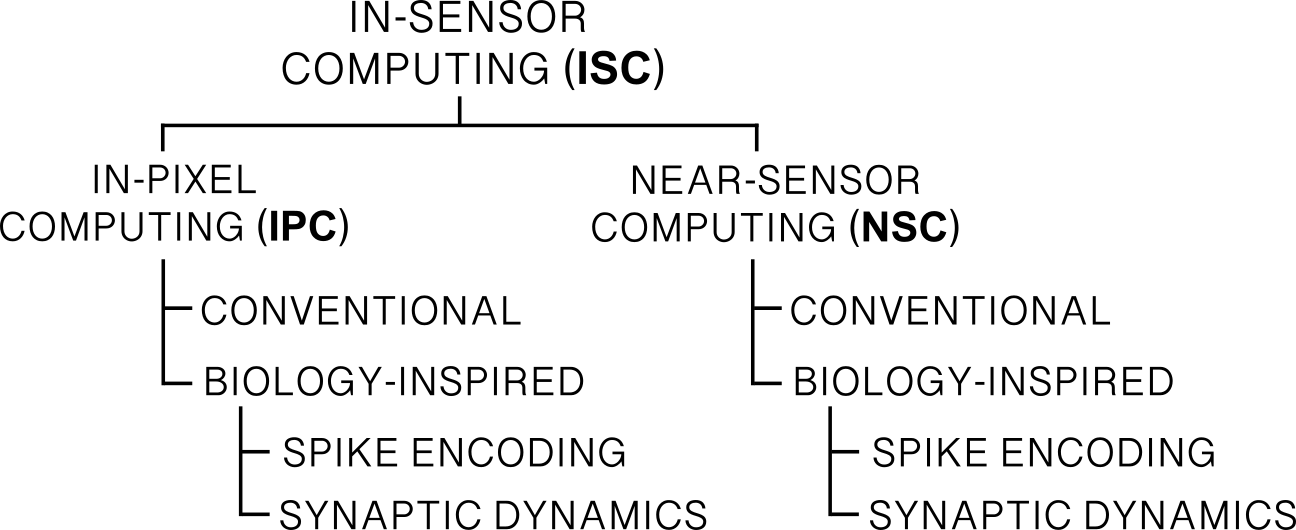}
    \caption{\textbf{Approaches to Intelligent Sensing.} When computation is integrated directly within the physical sensor unit, the broader paradigm of in-sensor computing is realized. ISC can be classified based on where computation occurs: either within individual pixels (in-pixel computing), or in processing units tightly coupled to the sensing layer (near-sensor computing). A second level of classification concerns how computation is implemented. In conventional approaches, it typically relies on digital processing; for example, embedding digitization at the pixel level in IPC, or using dedicated digital processing blocks in NSC. In contrast, emerging approaches are inspired by biological systems, including neuromorphic spike-based computation at the IPC level and in-memory computing paradigms at the NSC level, respectively.}
    \label{fig:1}
\end{wrapfigure}

\noindent Advances in science have historically been driven by the development of instruments that extend human sensory capabilities. Over time, these tools have evolved into the sophisticated yet ubiquitous devices we now call sensors. Early sensors, such as thermometers, mercury barometers, and magnetic compasses, were standalone mechanical or chemical systems. They did not need external power or computation. With the rise of electronic computing, sensors became the perceptual front ends of digital systems. Today, sensors are no longer seen as passive recorders of physical signals. They can be part of computational systems or stand alone and respond intelligently to their environment\cite{hepp2026mapping}.  This transformation is unfolding largely along architectural paths. The established approach uses conventional digital microarchitecture, integrating \emph{microcontroller units}\cite{ahmmed2019wearable,lee2018low,st2024stm32h7,infineon2023aurix,renesas2023rh850} (MCUs) or \emph{microprocessor units}\cite{openai2024gpt4o,TESLAFELETAI,tesla2023fleetai,NVIDIA_Orin_HotChips,apple2024visionpro,Gaoetal2024}(MPUs) directly within sensing hardware, but separated from the sensor unit. These systems follow traditional Von Neumann principles, keeping sensing, memory, and computation separate, and rely on synchronous, clock-driven operation, ultimately being limited by CMOS process constraints and interconnects. A more recent and disruptive approach integrates computation directly into the physical sensor unit (see Figure \ref{fig:1}). In this \emph{in-sensor computing} (ISC) paradigm, computation is performed either within the sensing element itself at the pixel level (\emph{in-pixel computing} (IPC)) or in close physical proximity to the sensing array (\emph{near-sensor computing} (NSC)). Emerging implementations of these architectures draw inspiration from biological systems, particularly the mammalian brain\cite{SonyIMX500_ISSCC,bonazzi2023tinytrackerultrafastultralowpoweredge,zhou20258,Prophesee_ISSCC2024,synsense2025speck,brainchip2025akida,Cimarellietal2025,richter2023speck,Pedersenetal2023}. This paradigm is evolving along two complementary directions. The first emulates biological computation through spike-based encoding, where information is transmitted only upon changes in the input signal, broadly referred to as a \emph{neuromorphic} architecture. The second focuses on structural integration, in which sensing and computation are physically co-located and tightly coupled through synaptic dynamics\cite{mennel2020ultrafast}, broadly referred to as \emph{in-memory computing} (IMC).

In the first of these biology inspired approaches, sensing elements can approximate the continuous differential equations governing neuronal behavior directly in the physics of the underlying circuits. In this neuromorphic architecture, each sensory pixel acts as an independent unit, similar to a biological neuron. It integrates the input signal and fires in response to temporal changes in the stimulus. The output is encoded as streams of spike trains, address event representations, or a combination of both. These representations are generally described as spike based encoding. A prominent example of a neuromorphic framework is event based neuromorphic vision sensors. Unlike traditional frame based vision sensors, which sample light intensity at fixed intervals, event-based neuromorphic sensors respond asynchronously to logarithmic changes in illumination\cite{chen2020event}. This spike driven principle extends to other sensory modalities, including auditory, tactile, and olfactory\cite{wu2024spatiotemporal,gao2025neuromorphic,kang2025neuromorphic}. In certain neuromorphic architectures, the physical structure of the sensory pixel becomes morphologically similar to neuronal motifs\cite{mahowald1994silicon,kim2018bioinspired}. The pixel embeds not only neuronal function but also structural features that resemble biological counterparts. The second approach enables computation directly within the sensor. In this architecture, the sensing elements can remain conventional. However, instead of simply streaming raw data off sensor, processing can take place locally. This localization can be achieved either through tight integration with a processing unit or by enabling computational capabilities within individual pixels. A prominent example of the first approach here are vision systems that stack an image processing digital tier with the sensing array. In this scheme, pixel data can be stored and processed in real time within the processing tier\cite{nakamura2024novel}. For spike-based sensor pixels, the processing tier itself can adopt a neuromorphic architecture, leveraging rich neuronal dynamics such as spiking neural networks. In the second approach, certain computational capabilities are embedded at the pixel level. This allows the sensor array to produce preprocessed outputs instead of raw signals. These outputs can then be further processed in a processing unit for more advanced operations\cite{syed2025phase,zolfagharinejad2025analogue}.\\


\begin{figure}[h!]
    \includegraphics[width=1\textwidth]{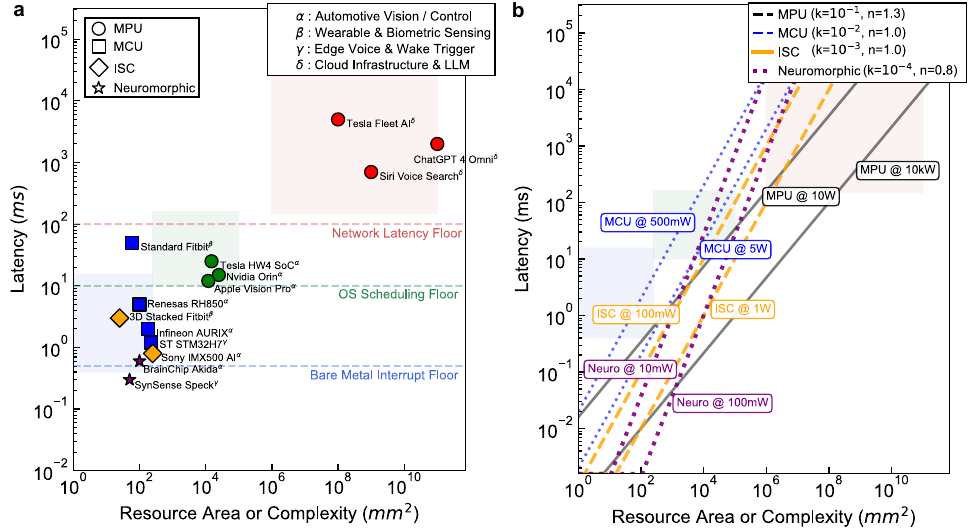}
    \caption{\textbf{Power Delay Area Mapping Framework.} (a) The plot categorizes sensors according to their resource/compute complexity and reaction latency. The values shown represent approximate estimates intended to illustrate general trends. The dotted horizontal lines represent the fundamental latency floors associated with different architectural paradigms. As the characteristic distance scales down from millimeters to microns, these latency floors shift from traditional MPU-based systems toward near sensor and neuromorphic architectures, reflecting the architectural transitions enabled by tighter integration. Note that not all data points correspond to hardware designed for workload flexibility or complex computations, and therefore vary in compute complexity. For example, all MCU data points are primarily optimized for fixed-function control tasks. (b) Iso-contours of power consumption derived from the PDA mapping delineate the different architectural regimes. These contours reveal the underlying efficiency limits and associated constraints.}
    \label{fig:2}
\end{figure}

This growing diversity of approaches to embedding intelligence in sensors raises important questions about the future ecosystem and scaling trajectories. The strong asymmetric resource constraints across different sensing modalities and applications, however make it difficult to develop a unified framework. For example, automotive platforms can tolerate large resource budgets, such as chip area and thermal design power, but demand very low latency. In contrast, wearable biometric sensors operate under strict power and area limits but can tolerate higher latency. We develop a Power-Delay-Area (PDA) mapping framework to address this (see Figure \ref{fig:2}a-b). At the heart of the framework is a semi-empirical scaling relationship $P = k \cdot \frac{A}{L^{n}}$, where $P$ denotes power, $A$ denotes the functional/active area (hardware complexity), and $L$ latency. This formulation encodes three key insights. First, the constant $k$ defines the intrinsic 
energy floor of an architecture, determined by the physical distance of data movement, compute organization and memory hierarchy. Second, the ratio $A/L$ reflects the commonly noted reality that increasing system complexity (large $A$) or demanding rapid response (small $L$) inherently increases power consumption. Third, the exponent $n$ characterizes the scaling efficiency of the architecture. While $k$ sets the 
magnitude of power consumption, $n$ dictates the penalty of performance; it captures 
how communication overhead and synchronization costs grow as latency is reduced and follows super--linear, linear, or sub--linear trends. 

In synchronous MPU architectures that utilize operating systems (OS) to enable multi-tasking, the scaling regime is typically super-linear ($n > 1$) due to the compounding costs of global coordination. To meet stringent timing constraints, interconnect topologies must adopt wider metal pitches, deeper pipelines, and aggressive buffering via repeaters and synchronization registers. These changes create a parasitic feedback loop: wider wires increase switching capacitance, while deeper pipelines introduce register overhead and branch-misprediction penalties. When latency is pushed further, the necessity of scaling supply voltage causes the power trajectory to further increase due to the quadratic relationship in the voltage-limited regimes. Simultaneously, the proportionality constant $k$ is inflated by the fixed infrastructure overhead. The inclusion of repeater chains, global clock trees to synchronize deep pipelines, and the leakage from additional synchronization registers significantly raises the architecture’s energy floor. $k$ is further inflated by data management overhead such as of coherence directories for cache management and the logic required for complex instruction-set decoding. Consequently, in these general-purpose regimes, the total power cost increases faster than the latency benefit, driven more by the systemic overhead of the physical cost of the communication fabric and data management.

MCU provides the first structural inflection point by enabling compute to be placed inside or immediately adjacent to the sensor, such as in system in package integration. This transition shifts the system toward a near–linear trajectory $(n \approx 1)$ because the dominant I/O energy and complexity associated with OS-level overhead, memory management, and pipelining are reduced, and clock frequency is typically fixed.  However, while the slope parameter $n$ improves, the baseline constant $k$ remains comparatively elevated because signals must still traverse long horizontal metal interconnects across the two-dimensional plane of the die. These lateral on-chip communication require high-current off-chip drivers and retain parasitic capacitance.  Three-dimensional stacking represents the next major step, leading to improved NSC, where sensing and computation are integrated within a single vertically organized structure. In this architecture, computational circuitry is placed directly beneath the pixel array and interconnected through dense vertical links such as through-silicon vias or hybrid bonds\cite{EkiSony,nakamura2024novel}. This vertical organization fundamentally changes the geometry of data movement by replacing long planar communication paths with extremely short vertical connections, bringing processing physically closer to where data is generated. Vertical integration collapses the physical distance between sensing and computation from millimeter-scale planar routes to micrometer-scale vertical interconnects, and ultimately to sub-micron copper-to-copper bonds. Because interconnect capacitance scales approximately with length, this short-wire topology directly reduces parasitic capacitance and lowers the baseline energy required for each operation ($k\downarrow$). The scaling exponent $n$ improves as 3D integration enables a transition from serialized planar communication to highly parallel data movement. As the pitch of vertical interconnects shrinks, their density increases quadratically, allowing many more signals to be transferred simultaneously between sensing and compute layers. This massive increase in parallel bandwidth reduces communication bottlenecks and drives the effective scaling behavior toward a more favorable, near-linear regime. Neuromorphic architectures represent the most energy-efficient operating regime, combining a sub-linear scaling exponent $(n < 1)$ with a small prefactor $k$. As spike encoding is implemented directly within the sensory pixels, neuromorphic architectures naturally realize the IPC framework. The sub-linear behavior originates from abandoning the global clock in favor of temporally sparse, event-driven and in some cases logarithmically compressed computation\cite{indiveri2019importance}. Because only active sensory pixels consume dynamic power, the penalty for reducing latency is confined to the pathways that are actually used, while the large inactive regions contribute essentially no marginal cost. Similarly, the prefactor $k$ is minimized through near-threshold or sub-threshold voltage operation and the use of low-leakage devices. In this regime, the interconnect component within $k$ is further optimized through localized peer-to-peer signaling, which avoids the overhead associated with global bus structures. In a 3D configuration, where sensing and computation are vertically co-located, $k$ approaches its minimal value. 

\begin{figure}[h!]
    \includegraphics[width=1\textwidth]{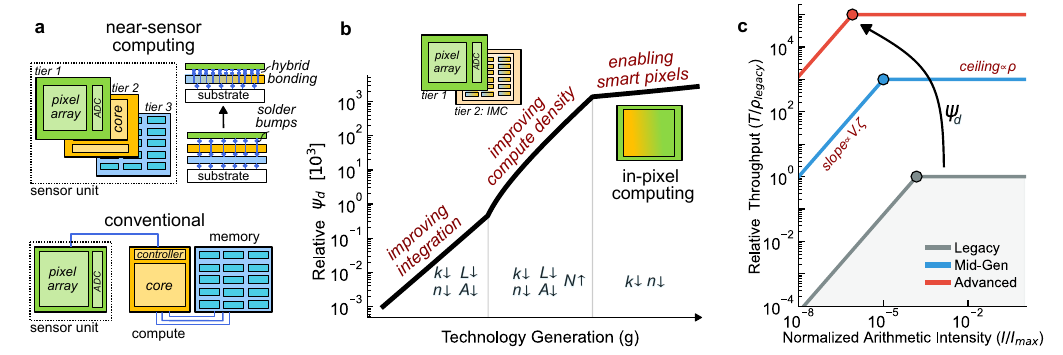}
    \caption{\textbf{Evolution of Architectures.} (a) Illustration of sensor architectures. Modern sensors are evolving from weakly interconnected 2D topologies (conventional) toward NSC, enabled by 3D integration of sensing, compute, and memory units. Further improvements in NSC arise from increased interconnect density through hybrid bonding. (b) Conceptual illustration highlighting the key dimensions driving this hardware evolution. As integration advances toward tighter interfacing, latency and data-movement energy decrease, also enabling progressively smaller form factors. These benefits are improved by higher compute density through increased number of processing elements and weight stationarity in IMC. As these approaches reach maturity, further performance improvements arise from IPC, where pixels perform sensing and computation locally. Additional gains, for specific tasks such as data recording, can be achieved with stateful dynamical pixels, which incorporate local memory elements within the pixel. (c) The combined increase in compute density and data-movement bandwidth enables sensors to move beyond memory-bound operation, even for low-arithmetic tasks. These are favorably gained while increasing computational throughput.}
    \label{fig:3}
\end{figure}

In this phenomenological perspective, architectural efficiency is not a single scalar metric but rather a coupled description of where the power curve originates and how steeply it rises under increasing performance pressure, represented by the parameters $(k,n)$. Within this framework, we define a metric $\psi$ that quantifies the physical efficiency with which an intelligent sensor converts sensed data into actionable computation. $\psi$ is a product of the architectural quality ($\zeta = \frac{1}{k \cdot n}$), as the intrinsic efficiency of the micro-architecture, scaled by the compute density ($\rho= \frac{T}{A}$), where $T$ denotes computational throughput, and scales directly with number of processing elements ($N$) and low latency. Current integration trends in sensing and emerging compute paradigms indicate a progressively increasing value of $\psi$. From an integration perspective (see Figure \ref{fig:3}a-b), as previously discussed, shorter interconnect lengths, result in lower energy per transfer and higher interconnect densities, which enable increased data transfer rates. From a computational perspective, shorter interconnect distances reduce latency. They also enable scaling to a large $N$, particularly through emerging paradigms such as IMC architectures. Moreover, the higher achievable data rates can allow processors to be overclocked, enabling higher throughput. Using $\psi$, the evolution of sensor generations can be projected as they become progressively optimized for architecture and compute. This progression is described in three distinct phases, each characterized by the dominant source of improvement. The initial phase is primarily driven by enhancements in $\zeta$. The second phase is characterized by improvements in $\rho$, enabled by scaling up the number of processing elements ($N$) and by architectural strategies such as weight stationarity through IMC, and close-memory access through 3D-integration. The final phase transitions to an event-driven regime, in which performance gains are again predominantly governed by advances in $\zeta$. These improvements can be further casted onto roofline framework by defining the parameter $\psi$ as a measure of how closely a system can approach peak throughput across a wide range of arithmetic intensities. Increasing $\psi$ can enable compute near the peak-performance limit even for low algorithmic arithmetic intensity workloads, thereby making the sensor a favorable compute-bound hardware platform rather than a bandwidth-limited one (see Figure \ref{fig:3}c).

Even further gains in sensor efficiency become possible when computation and memory are embedded directly within the sensory pixel. In this \emph{in-pixel-in-memory} (IPC-M) framework regime, $\psi$ approaches a theoretical asymptote, representing its fundamental upper bound, as architectural and interconnect overheads are minimized through the statefulness of the system. Within such architectures, $k$ transitions from representing an interconnect penalty to capturing a direct physical state transformation, while $n$ converges toward its most efficient event-driven limit. The sensory pixels are realized using emerging materials capable of programmable in-pixel computation during sensing\cite{syed2025phase}, or through integration with memory elements that enable reconfigurable operation\cite{zhou2019optoelectronic}. In both cases, the pixel output becomes a precomputed value rather than raw data\cite{syed2025phase}. For single-stage computational tasks, such as image recording or filtering, this approach approaches maximal efficiency. For more complex, multi-stage processing pipelines, it can serve as an effective front-end stage, for example as the initial layer of a convolutional neural network, that can be combined with the 3D NSC. This remains an active and evolving research area, with ongoing efforts aimed at incorporating temporal dynamics directly within the pixel, including short-term plasticity and related adaptive mechanisms\cite{chen2023optoelectronic,xu2025high,zhang2025emerging,sarwat2022phase}.

In summary, we have outlined and categorized the evolving landscape of sensor technologies and the technological advances shaping their future. Crucially, sensors are improving not only in their ability to capture signals with greater fidelity, but also in their ability to process sensed data more efficiently. This dual progression (enhancing both sensing and computation) marks a broader transition from the era of transistor density, toward an era defined by intelligence density, where the value of a system will be increasingly determined by how effectively it can transform sensed information into actionable computation. This, in turn, will be governed by how efficiently data can be communicated, underscoring progress trends in interconnects, compute and system integration.

\newpage

\section*{Methods}

\paragraph{Area–Latency Scaling:} Latency can be considered in a normalized form as $L/L_{\mathrm{ref}}$, where $L_{\mathrm{ref}} = 1\,\mathrm{s}$ is chosen to fix the units of $k$. On a log--log scale, the PDA analysis yields a first-order linear relationship, $
\log L = \frac{1}{n}\log A + \frac{1}{n}\log\!\left(\frac{k}{P}\right)$. The slope $m = \tfrac{1}{n}$ captures area–latency sensitivity, while the intercept $b = \tfrac{1}{n}\log\!\left(\tfrac{k}{P}\right)$ reflects the influence of architectural efficiency $k$. Architectures with smaller $n$ benefit more strongly from area scaling, whereas reductions in $k$ shift the curve downward.

\paragraph{Intelligence Density Metric:} On the account of the used approach, we simplify the definition of intelligence to the capacity for high-efficiency information processing. While we discount the qualitative nature or breadth of the processing (reasoning, logic, etc.), this allows for a deterministic metric, $\psi$, which measures the physical limits of hardware in converting sensed signals into actionable compute. The theoretical scaling of $\psi$ follows the relationship $\psi(g) = \left[ \frac{1}{k(g)\, n(g)} \right]\left[ \frac{N(g)}{A(g)\, L(g)} \right]$, where $g$ denotes the technology generation. The theoretical frontier can be interpreted as the sum of the individual scaling contributions. Expressing the relationship in logarithmic form, and for a select generation yields $
\log \psi = \log N-\log\!\left(k\right) -\log\!\left(n\right)  -\log A-\log L$. This expression shows that the scaling trajectory of $\psi$ is determined by the additive contributions of improvements in architectural efficiency $(k,n)$ and compute density $(N/LA)$. Within a roofline-style analysis, $\zeta$ should be interpreted primarily as a measure of the energy efficiency of data movement, rather than bandwidth itself. Bandwidth and $\zeta$ become directly related only under the constraint of a fixed power budget. In this regime, higher energy efficiency in data movement enables proportionally higher achievable $\text{bandwidth} = V \times \frac{\text{Power}}{\text{Energy per Byte}}$, where $V$ represents the number of data lanes/vias. The peak performance, however, is a direct function of the compute density ($\rho$). The knee point is defined as the arithmetic intensity at which bandwidth equals peak performance, given by $I_{\text{knee}} = \frac{\rho}{\zeta}$.

\section*{COI Statement}
\noindent The author declare no competing interests.

\section*{Acknowledgments}
\noindent The author would like to thank the MemVerse group (Dr. Abu Sebastian, Dr. Manuel Le Gallo, Mr. Loris Coccia, Ms. Jinane Bazzi) at IBM Research Zurich for their proofreading of the manuscript. The work is supported through European Research Council Starting Grant INFUSED (101222715).
\section*{References}
\bibliographystyle{naturemag}
\bibliography{References}

@misc{hepp2026mapping,
  title        = {Mapping the Automotive Software and Electronics Landscape},
  author       = {Dominik Hepp and Martin Kellner and Sören Jautelat and Michael Guggenheimer and Tomás Aloise},
  year         = {2026},
  month        = {January 6},
  howpublished = {\url{https://www.mckinsey.com/features/mckinsey-center-for-future-mobility/our-insights/mapping-the-automotive-software-and-electronics-landscape}}
}

@article{syed2025phase,
  title={Phase change computational sensor},
  author={Syed, Ghazi Sarwat and Kersting, Benedikt and Egger, Urs and Sebastian, Abu},
  journal={npj Unconventional Computing},
  volume={2},
  number={1},
  pages={1},
  year={2025},
  publisher={Nature Publishing Group UK London}
}

@article{chen2023optoelectronic,
  title={Optoelectronic graded neurons for bioinspired in-sensor motion perception},
  author={Chen, Jiewei and Zhou, Zheng and Kim, Beom Jin and Zhou, Yue and Wang, Zhaoqing and Wan, Tianqing and Yan, Jianmin and Kang, Jinfeng and Ahn, Jong-Hyun and Chai, Yang},
  journal={Nature Nanotechnology},
  volume={18},
  number={8},
  pages={882--888},
  year={2023},
  publisher={Nature Publishing Group UK London}
}

@inproceedings{zhou20258,
  title={An 8.62 $\mu$W 75dB-DR SoC End-to-End Spoken-Language-Understanding SoC With Channel-Level AGC and Temporal-Sparsity-Aware Streaming-Mode RNN},
  author={Zhou, Sheng and Li, Zixiao and Delbruck, Tobi and Kim, Kwantae and Liu, Shih-Chii},
  booktitle={2025 IEEE International Solid-State Circuits Conference (ISSCC)},
  volume={68},
  pages={238--240},
  year={2025},
  organization={IEEE}
}

@inproceedings{nakamura2024novel,
  title={A novel 1/1.3-inch 50 megapixel three-wafer-stacked cmos image sensor with dnn circuit for edge processing},
  author={Nakamura, R and Tsugawa, H and Yamagishi, H and Fujisaki, Y and Suda, Y and Tatsumi, Y and Shimizu, K and Kagawa, Y and Ono, K and Horie, Y and others},
  booktitle={2024 IEEE International Electron Devices Meeting (IEDM)},
  pages={1--4},
  year={2024},
  organization={IEEE}
}

@INPROCEEDINGS{EkiSony,
  author={Eki, Ryoji and Yamada, Satoshi and Ozawa, Hiroyuki and Kai, Hitoshi and Okuike, Kazuyuki and Gowtham, Hareesh and Nakanishi, Hidetomo and Almog, Edan and Livne, Yoel and Yuval, Gadi and Zyss, Eli and Izawa, Takashi},
  booktitle={2021 IEEE International Solid-State Circuits Conference (ISSCC)}, 
  title={9.6 A 1/2.3inch 12.3Mpixel with On-Chip 4.97TOPS/W CNN Processor Back-Illuminated Stacked CMOS Image Sensor}, 
  year={2021},
  volume={64},
  number={},
  pages={154-156},
  doi={10.1109/ISSCC42613.2021.9365965}}

@article{zolfagharinejad2025analogue,
  title={Analogue speech recognition based on physical computing},
  author={Zolfagharinejad, Mohamadreza and B{\"u}chel, Julian and Cassola, Lorenzo and Kinge, Sachin and Syed, Ghazi Sarwat and Sebastian, Abu and van der Wiel, Wilfred G},
  journal={Nature},
  volume={645},
  number={8082},
  pages={886--892},
  year={2025},
  publisher={Nature Publishing Group UK London}
}

@article{chen2020event,
  title={Event-based neuromorphic vision for autonomous driving: A paradigm shift for bio-inspired visual sensing and perception},
  author={Chen, Guang and Cao, Hu and Conradt, Jorg and Tang, Huajin and Rohrbein, Florian and Knoll, Alois},
  journal={IEEE Signal Processing Magazine},
  volume={37},
  number={4},
  pages={34--49},
  year={2020},
  publisher={IEEE}
}

@article{indiveri2019importance,
  title={The importance of space and time for signal processing in neuromorphic agents: the challenge of developing low-power, autonomous agents that interact with the environment},
  author={Indiveri, Giacomo and Sandamirskaya, Yulia},
  journal={IEEE Signal Processing Magazine},
  volume={36},
  number={6},
  pages={16--28},
  year={2019},
  publisher={IEEE}
}

@online{openai2024gpt4o,
  author = {OpenAI},
  title = {Hello GPT-4o},
  year = {2024},
  url = {https://openai.com/index/hello-gpt-4o/},
}

@article{tesla2023fleetai,
  author = {Tesla},
  title = {Tesla AI Day 2022 Proceedings},
  journal = {Tesla Investor Relations},
  url = {https://www.youtube.com/watch?v=ODSJsviD_SU},
  year = {2023},
}

@ARTICLE{TESLAFELETAI,
  author={Talpes, Emil and Sarma, Debjit Das and Williams, Doug and Arora, Sahil and Kunjan, Thomas and Floering, Benjamin and Jalote, Ankit and Hsiong, Christopher and Poorna, Chandrasekhar and Samant, Vaidehi and Sicilia, John and Nivarti, Anantha Kumar and Ramachandran, Raghuvir and Fischer, Tim and Herzberg, Ben and McGee, Bill and Venkataramanan, Ganesh and Banon, Pete},
  journal={IEEE Micro}, 
  title={The Microarchitecture of DOJO, Tesla’s Exa-Scale Computer}, 
  year={2023},
  volume={43},
  number={3},
  pages={31-39},
  doi={10.1109/MM.2023.3258906}}

@online{apple2024visionpro,
  author = {Apple Inc.},
  title = {Apple Vision Pro Technical Specifications},
  year = {2024},
  url = {https://www.apple.com/apple-vision-pro/specs/}
}

@manual{st2024stm32h7,
  title = {STM32H743xI/G High-performance MCU Datasheet},
  organization = {STMicroelectronics},
  year = {2024},
}

@manual{infineon2023aurix,
  title = {AURIX TC3Ex Product Brief},
  organization = {Infineon Technologies AG},
  year = {2023},
}

@manual{renesas2023rh850,
  title = {RH850/F1KM-S1 Automotive Microcontroller},
  organization = {Renesas Electronics},
  year = {2023},
}

@misc{bonazzi2023tinytrackerultrafastultralowpoweredge,
      title={TinyTracker: Ultra-Fast and Ultra-Low-Power Edge Vision In-Sensor for Gaze Estimation}, 
      author={Pietro Bonazzi and Thomas Ruegg and Sizhen Bian and Yawei Li and Michele Magno},
      year={2023},
      eprint={2307.07813},
      archivePrefix={arXiv},
      primaryClass={cs.CV},
      url={https://arxiv.org/abs/2307.07813}, 
}

@manual{synsense2025speck,
  title = {Speck: Integrated Neuromorphic Vision SoC User Guide},
  organization = {SynSense},
  year = {2025},
}

@manual{brainchip2025akida,
  title = {Akida AKD1000 System-on-Chip Product Brief},
  organization = {BrainChip Holdings Ltd.},
  year = {2025},
}

@article{NVIDIA_Orin_HotChips,
  author  = {Dally, Bill and et al.},
  title   = {The NVIDIA Orin Architecture: A 275 TOPS System-on-Chip for Autonomous Vehicles},
  journal = {Hot Chips 34 Symposium},
  year    = {2022},
  note    = {Detailed architectural specs for 256 Tensor Cores and Ampere GPU clusters}
}

@article{Gaoetal2024,
  author  = {Gao, Peng and Liu, Yang and Wang, Jun and Cai, Wanlin and Shen, Guangchong and Hong, Zonghui and Qu, Jiali and Wang, Ning},
  title   = {SOPHGO BM1684X: A Commercial High Performance Terminal AI Processor with Large Model Support},
  journal = {2024 57th IEEE/ACM International Symposium on Microarchitecture (MICRO)},
  year    = {2024},
  pages   = {1413--1428},
  doi     = {10.1109/micro61859.2024.00104}
}

@INPROCEEDINGS{SonyIMX500_ISSCC,
  author={Eki, Ryoji and Yamada, Satoshi and Ozawa, Hiroyuki and Kai, Hitoshi and Okuike, Kazuyuki and Gowtham, Hareesh and Nakanishi, Hidetomo and Almog, Edan and Livne, Yoel and Yuval, Gadi and Zyss, Eli and Izawa, Takashi},
  booktitle={2021 IEEE International Solid-State Circuits Conference (ISSCC)}, 
  title={9.6 A 1/2.3inch 12.3Mpixel with On-Chip 4.97TOPS/W CNN Processor Back-Illuminated Stacked CMOS Image Sensor}, 
  year={2021},
  volume={64},
  number={},
  pages={154-156},
  doi={10.1109/ISSCC42613.2021.9365965}}

@article{Prophesee_ISSCC2024,
  author  = {Finateu, Thomas and et al.},
  title   = {A 320x320 Back-Illuminated Stacked Event-Based Vision Sensor with 4.88um Pixel Pitch and 150dB Dynamic Range},
  journal = {IEEE International Solid-State Circuits Conference (ISSCC)},
  year    = {2024},
  note    = {Primary reference for GenX320 low-latency event-based sensing}
}

@article{Cimarellietal2025,
  author  = {Cimarelli, Claudio and Millan-Romera, Jose Andres and Voos, Holger and Sanchez-Lopez, Jose Luis},
  title   = {Hardware, Algorithms, and Applications of the Neuromorphic Vision Sensor: a Review},
  journal = {arXiv},
  year    = {2025},
  doi     = {10.48550/arxiv.2504.08588}
}

@article{Pedersenetal2023,
  author  = {Pedersen, Jens E. and Abreu, Steven and Jobst, Matthias and Lenz, Gregor and Fra, Vittorio and Bauer, Felix C. and Muir, Dylan R. and Zhou, Peng and Vogginger, Bernhard and Heckel, Kade and Urgese, Gianvito and Shankar, Sadasivan and Stewart, Terrence C. and Sheik, Sadique and Eshraghian, Jason K.},
  title   = {Neuromorphic Intermediate Representation: A Unified Instruction Set for Interoperable Brain-Inspired Computing},
  journal = {Nat Commun 15, 8122 (2024)},
  year    = {2023},
  doi     = {10.1038/s41467-024-52259-9}
}

@article{richter2023speck,
  title={Speck: A smart event-based vision sensor with a low latency 327k neuron convolutional neuronal network processing pipeline},
  author={Richter, Ole and Xing, Yannan and De Marchi, Michele and Nielsen, Carsten and Katsimpris, Merkourios and Cattaneo, Roberto and Ren, Yudi and Hu, Yalun and Liu, Qian and Sheik, Sadique and others},
  journal={arXiv preprint arXiv:2304.06793},
  year={2023}
}

@article{lee2018low,
  title={A low-power photoplethysmogram-based heart rate sensor using heartbeat locked loop},
  author={Lee, Jinseok and Jang, Do-Hun and Park, Sujin and Cho, SeongHwan},
  journal={IEEE transactions on biomedical circuits and systems},
  volume={12},
  number={6},
  pages={1220--1229},
  year={2018},
  publisher={IEEE}
}

@inproceedings{ahmmed2019wearable,
  title={A wearable wrist-band with compressive sensing based ultra-low power photoplethysmography readout circuit},
  author={Ahmmed, Parvez and Dieffenderfer, James and Valero-Sarmiento, Jose Manuel and Pamula, Venkata Rajesh and Van Helleputte, Nick and Van Hoof, Chris and Verhelst, Marian and Bozkurt, Alper},
  booktitle={2019 IEEE 16th international conference on wearable and implantable body sensor networks (BSN)},
  pages={1--4},
  year={2019},
  organization={IEEE}
}

@article{sarwat2022phase,
  title={Phase-change memtransistive synapses for mixed-plasticity neural computations},
  author={Sarwat, Syed Ghazi and Kersting, Benedikt and Moraitis, Timoleon and Jonnalagadda, Vara Prasad and Sebastian, Abu},
  journal={Nature Nanotechnology},
  volume={17},
  number={5},
  pages={507--513},
  year={2022},
  publisher={Nature Publishing Group UK London}
}

@article{zhang2025emerging,
  title={Emerging materials and computing paradigms for temporal signal analysis},
  author={Zhang, Teng and Wozniak, Stanislaw and Syed, Ghazi Sarwat and Mannocci, Piergiulio and Farronato, Matteo and Ielmini, Daniele and Sebastian, Abu and Yang, Yuchao},
  journal={Advanced Materials},
  volume={37},
  number={12},
  pages={2408566},
  year={2025},
  publisher={Wiley Online Library}
}

@article{gao2025neuromorphic,
  title={A neuromorphic robotic electronic skin with active pain and injury perception},
  author={Gao, Yuyu and Zhang, Jianpeng and Zhang, Hehua and Chow, Lung and Zhao, Guangyao and Guo, Guihuan and Yiu, Chun Ki and Zhang, Binbin and Huang, Ya and Zhou, Jingkun and others},
  journal={Proceedings of the National Academy of Sciences},
  volume={122},
  number={52},
  pages={e2520922122},
  year={2025},
  publisher={National Academy of Sciences}
}

@article{kang2025neuromorphic,
  title={Neuromorphic olfaction with ultralow-power gas sensors and ovonic threshold switch},
  author={Kang, Mingu and Han, Joon-Kyu and Lee, Kichul and Jeong, Jaeseok and Yoo, Chanyoung and Jeon, Jeong Woo and Park, Byongwoo and Choi, Wonho and Ahn, Junseong and Yoon, Kuk-Jin and others},
  journal={Science Advances},
  volume={11},
  number={39},
  pages={eadv9222},
  year={2025},
  publisher={American Association for the Advancement of Science}
}

@article{wu2024spatiotemporal,
  title={Spatiotemporal audio feature extraction with dynamic memristor-based time-surface neurons},
  author={Wu, Xulei and Dang, Bingjie and Zhang, Teng and Wu, Xiulong and Yang, Yuchao},
  journal={Science Advances},
  volume={10},
  number={14},
  pages={eadl2767},
  year={2024},
  publisher={American Association for the Advancement of Science}
}

@article{kim2018bioinspired,
  title={A bioinspired flexible organic artificial afferent nerve},
  author={Kim, Yeongin and Chortos, Alex and Xu, Wentao and Liu, Yuxin and Oh, Jin Young and Son, Donghee and Kang, Jiheong and Foudeh, Amir M and Zhu, Chenxin and Lee, Yeongjun and others},
  journal={Science},
  volume={360},
  number={6392},
  pages={998--1003},
  year={2018},
  publisher={American Association for the Advancement of Science}
}

@incollection{mahowald1994silicon,
  title={The silicon retina},
  author={Mahowald, Misha},
  booktitle={An Analog VLSI System for Stereoscopic Vision},
  pages={4--65},
  year={1994},
  publisher={Springer}
}

@article{xu2025high,
  title={High-order dynamics in an ultra-adaptive neuromorphic vision device},
  author={Xu, Jiayi and Jiang, Biyi and Wang, Weizhen and Guo, Zhifeng and Gao, Junsen and Hu, Xinyan and Qin, Jingkai and Ran, Liang and Lin, Longyang and Cai, Songhua and others},
  journal={Nature nanotechnology},
  volume={20},
  number={10},
  pages={1419--1430},
  year={2025},
  publisher={Nature Publishing Group UK London}
}

@article{zhou2019optoelectronic,
  title={Optoelectronic resistive random access memory for neuromorphic vision sensors},
  author={Zhou, Feichi and Zhou, Zheng and Chen, Jiewei and Choy, Tsz Hin and Wang, Jingli and Zhang, Ning and Lin, Ziyuan and Yu, Shimeng and Kang, Jinfeng and Wong, H-S Philip and others},
  journal={Nature nanotechnology},
  volume={14},
  number={8},
  pages={776--782},
  year={2019},
  publisher={Nature Publishing Group UK London}
}

@article{mennel2020ultrafast,
  title={Ultrafast machine vision with 2D material neural network image sensors},
  author={Mennel, Lukas and Symonowicz, Joanna and Wachter, Stefan and Polyushkin, Dmitry K and Molina-Mendoza, Aday J and Mueller, Thomas},
  journal={Nature},
  volume={579},
  number={7797},
  pages={62--66},
  year={2020},
  publisher={Nature Publishing Group UK London}
}


\end{document}